\documentclass[acmsmall,screen]{acmart}
\renewcommand\footnotetextcopyrightpermission[1]{} 
\renewcommand{\acmArticleType}{}
\AtBeginDocument{%
  }
\usepackage{algorithm}
\usepackage{algorithmic}

\usepackage{utfsym}

\usepackage{tabularx} 
\usepackage{array}

\usepackage{multirow} 
\usepackage{booktabs} 
\usepackage{makecell} 
\usepackage{array} 
\usepackage{multirow} 

\usepackage{ragged2e} 

\usepackage{caption}
\usepackage{array} 

\usepackage{array} 

\usepackage{colortbl} 
\definecolor{lightgray}{gray}{0.8} 

\usepackage[utf8]{inputenc}
\usepackage{listing}
\usepackage{xcolor}

\setcopyright{acmlicensed}
\copyrightyear{2025}
\acmYear{2025}
\acmDOI{XXXXXXX.XXXXXXX}

\begin{document}

\title{Comparative Analysis of Blockchain Systems}


\author{Jiaqi Huang}
\affiliation{
  \institution{Hainan University}
  \city{Haikou}
  \country{China}
}

\author{Yuanzheng Niu}
\affiliation{%
  \institution{Hainan University}
  \city{Haikou}
  \country{China}}
  \email{niunicole1@outlook.com}

\author{Xiaoqi Li}
\affiliation{%
  \institution{Hainan University}
  \city{Haikou}
  \country{China}}
\email{csxqli@ieee.org}

\author{Zongwei Li}
\affiliation{%
  \institution{Hainan University}
  \city{Haikou}
  \country{China}}
  \email{lizw1017@gmail.com}

\renewcommand{\shortauthors}{ }

\begin{abstract}
Blockchain is a type of decentralized distributed database. Unlike traditional relational database management systems, it does not require management or maintenance by a third party. All data management and update processes are open and transparent, solving the trust issues of centralized database management systems. Blockchain ensures network-wide consistency, consensus, traceability, and immutability. Under the premise of mutual distrust between nodes, blockchain technology integrates various technologies, such as P2P protocols, asymmetric encryption, consensus mechanisms, and chain structures. Data is distributed and stored across multiple nodes, maintained by all nodes, ensuring transaction data integrity, undeniability, and security. This facilitates trusted information sharing and supervision.
The basic principles of blockchain form the foundation for all related research. Understanding the working principles is essential for further study of blockchain technology. There are many platforms based on blockchain technology, and they differ from one another. This paper will analyze the architecture of blockchain systems at each layer, focusing on the principles and technologies of blockchain platforms such as Bitcoin, Ethereum, and Hyperledger Fabric. The analysis will cover their scalability and security and highlight their similarities, differences, advantages, and disadvantages.
\end{abstract}

\ccsdesc[500]{Information systems~Distributed databases}
\ccsdesc[300]{Security and privacy~Cryptography}
\ccsdesc[200]{Networks~Peer-to-peer protocols}
\ccsdesc[200]{Security and privacy~Distributed systems security}

\keywords{Blockchain, Bitcoin, Ethereum, Hyperledger Fabric}

\maketitle
\fancyfoot{}
\pagestyle{plain} 

\section{Introduction}
Bitcoin is the world’s first decentralized cryptocurrency proposed by Satoshi Nakamoto\cite{1}. Essentially, it is a digital currency generated by a distributed network system, and its issuance process does not rely on a central authority. Instead, it is governed collectively by all the nodes in the distributed network. As the underlying technology of Bitcoin, blockchain\cite{03} has gained widespread attention with Bitcoin's popularity.
Ethereum\cite{2}, proposed by Buterin, is the second most popular blockchain platform after Bitcoin. In addition to enabling decentralized digital currency transactions\cite{02}, it provides a Turing-complete programming language for writing smart contracts, marking the first application of smart contracts on the blockchain\cite{01}.
Hyperledger\cite{3}, an open-source blockchain project developed by the Linux Foundation, aims to create cross-industry business platforms. Due to varying needs and service requirements across different sectors, different blockchains need to be constructed. Hyperledger offers several blockchain projects, with Fabric being the most notable. Unlike Bitcoin and Ethereum, Hyperledger Fabric is specifically designed for enterprise-level blockchain applications and introduces a membership management service.

In addition to the three blockchain projects mentioned above, numerous other representative initiatives exist in the industry. This paper focuses on analyzing these three blockchain systems, as shown in Table 1. Section 1 presents the overall system architecture of blockchain systems; Section 2 compares blockchain data from three aspects: data structure, data model, and data storage; Section 3 briefly explains the differences in the network layers of the three blockchain systems; Section 4 provides a detailed analysis of the consensus mechanisms used in the three systems; Section 5 discusses smart contracts from three perspectives: programming languages, runtime environments, and operating principles; Section 6 briefly summarizes the main applications of the three systems; Section 7 analyzes the scalability solutions of the three systems; and Section 8 introduces solutions for data security and privacy.

\begin{table*}[h!]
\centering
\caption{Comparative Analysis of Different Blockchains}

\begin{tabular}{ p{3.2cm} p{3.2cm} p{3.2cm} p{3.2cm}}
\toprule
\textbf{Blockchain Platform} & \textbf{Bitcion} & \textbf{Etherum} & \textbf{Hyperledger Fabric}\\
\midrule
Entry Mechanism  & Public Chain & Public Chain  & Consortium Chain\\

Data Structure  & Merkle Tree / Blockchain Table& Merkle Patricia Tree / Blockchain Table  & Merkle Bocket Tree / Blockchain Table\\

Data Model  & 	Transaction-based Model & Account-based Model  & Account-based Model\\

Blockchain Storage & File Storage & Level DB  & File Storage \\

Network Layer  & TCP-based P2P & TCP-based P2P  & HTTP/2-based P2P\\

Consensus Layer  & POW & POW/POS & 	PBFT/SBFT\\

Programming Language  & Script-based & Solidity/Serpent  & 	Go/Java\\

Sandbox Environment  & - & EVM & Docker\\

Application Layer & Bitcoin Transactions & Dapp/Ethereum Transactions & Enterprise-level Blockchain Applications\\
\bottomrule
\end{tabular}%
\label{table1}%
\end{table*}

\section{System Architecture}
\begin{figure}[h!]
    \centering
    \includegraphics[height=0.2\textheight]{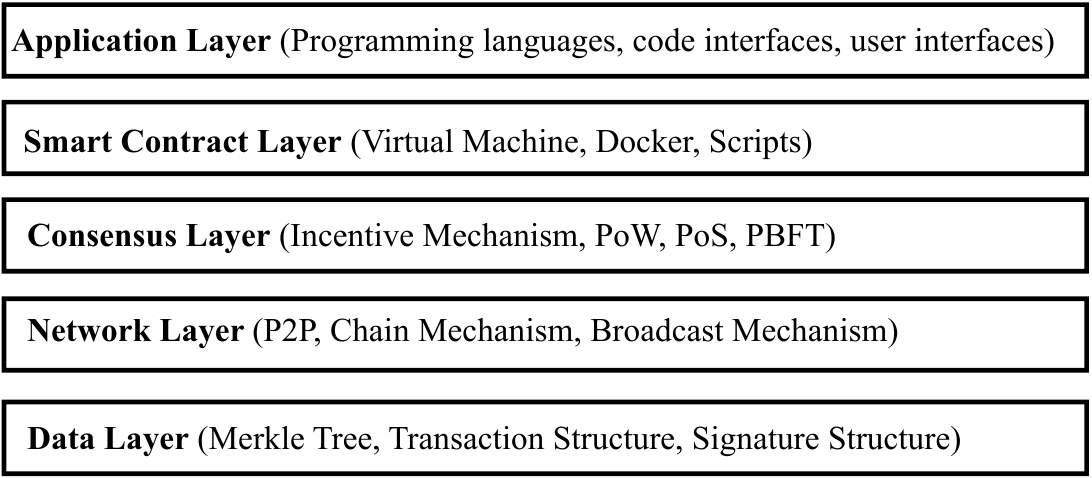}
    \caption{Blockchain System Architecture}
    \label{fig1}
\end{figure}

Many different platforms based on blockchain technology have emerged, each with its specific implementation, but there are many commonalities in their overall system architecture. As shown in Figure \ref{fig1}, blockchain platforms can generally be divided into five layers: from bottom to top, these are the Data Layer, Network Layer, Consensus Layer, Contract Layer, and Application Layer\cite{4}.

\section{Data Layer }
Blockchain technology\cite{06} utilizes two important concepts from cryptography\cite{5}: hash functions and signatures. Hash functions have three characteristics:

\begin{itemize}
\item {\texttt{Collision resistance}}:  It is impossible to find $m^{'} =m$ such that $H\left ( m^{'}  \right ) =H\left ( m \right ) $, meaning that it is not possible to tamper with data without detection.
\item{\texttt{Hiding}}: The process of calculating the hash value $H(x)$ for an input $x $is irreversible and can only be broken through brute force.
\item{\texttt{Puzzle friendly}}: The hash value computation is unpredictable in advance. One must try different possibilities without shortcuts to obtain the desired hash value.
\end{itemize}

\subsection{Data Structure}
\begin{figure}[h!]
    \centering
    \includegraphics[height=0.29\textheight]{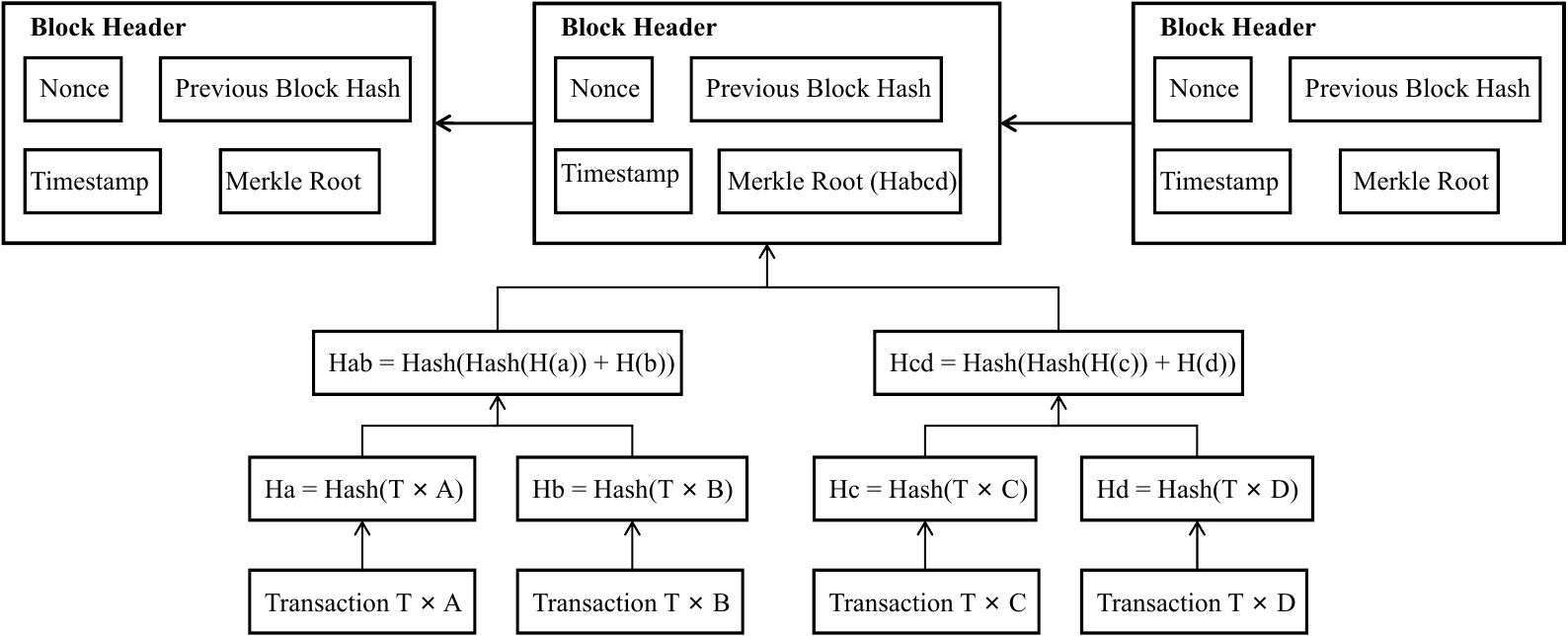}
    \caption{Blockchain Data Structure of the Bitcoin System}
    \label{fig2}
\end{figure}

Blockchain uses a linked list structure based on blocks and a Merkle tree to ensure data immutability. The blockchain data structure of the Bitcoin system is shown in Figure \ref{fig2}.

\subsubsection{Hash Pointer}
\begin{figure}[h!]
    \centering
    \includegraphics[height=0.1\textheight]{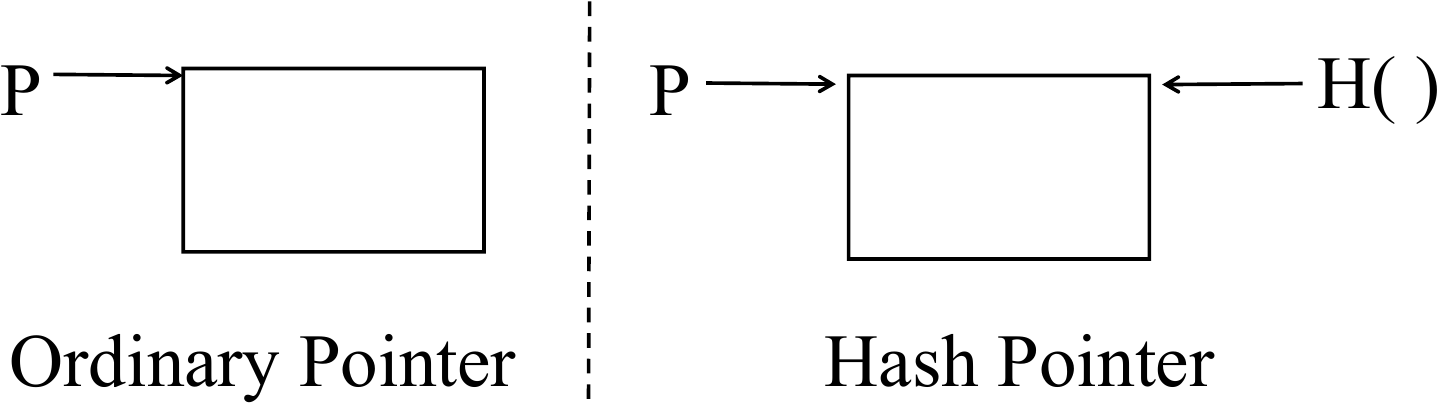}
    \caption{Ordinary pointers and Hash pointers}
    \label{fig 3}
\end{figure}

Blockchain uses hash pointers instead of ordinary pointers\cite{08}, as shown in Figure \ref{fig 3}. The content of the entire block is hashed together, creating a tamper-evident log. This means that if a value in any block is altered, the hash values of all subsequent blocks will also change, triggering a domino effect where a small change causes a chain reaction. If the hash value of the last block is known, it can be used to verify whether any previous block has been tampered with.

\subsubsection{Merkle Tree}
\begin{figure}
    \centering
    \includegraphics[width=0.7\linewidth]{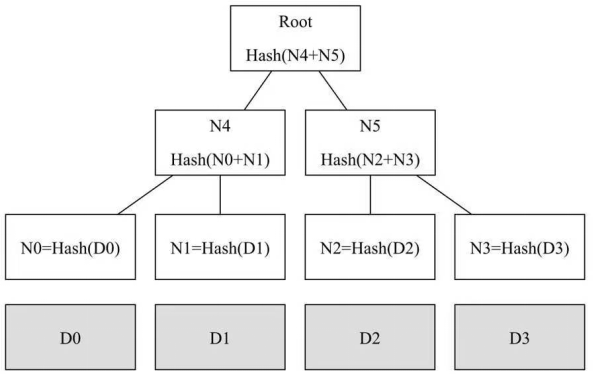}
    \caption{Mrekle tree}
    \label{fig4}
\end{figure}

Transactions and other activities in a block are hashed in the form of a Merkle tree, which is an important part of blockchain. It uses hash pointers instead of ordinary pointers, ensuring data authenticity, security, and non-repudiation. Bitcoin uses the simplest binary Merkle tree, as shown in Figure \ref{fig4}. Each transaction is hashed using SHA-256 and stored in the leaf nodes. The values of every two child nodes are connected, and then hashed to record them in the parent node. The final root hash of the transactions is stored in the Bitcoin block header. To confirm whether a specific transaction exists in the block, only the path from the transaction node to the Merkle root needs to be checked, without the involvement of other nodes in the tree. This is known as Simplified Payment Verification (SPV).
As shown in Figure 4, to verify whether transaction D1 exists in the block, one only needs to check if the hash of the transaction equals N1. If they are equal, N1 and N0 are concatenated and hashed, then compared with N4. If they match, the process repeats until it matches the Merkle root hash, confirming the existence of transaction D1 with a time complexity of $\Theta \left ( log\left ( N \right )  \right ) $. The entire Merkle tree can be transmitted to a lightweight node to prove that a transaction does not exist. If the tree's construction and the hash values at each level are correct, and no leaf node corresponds to the transaction, then the transaction does not exist, with a time complexity of $\Theta \left ( N \right ) $.

Ethereum uses the Merkle Patricia tree \cite{6}. The state data is extensive and frequently changes. When constructing a new block, only the changed account states need to be calculated, without recalculating the entire tree. The Merkle Patricia tree allows for fast lookup of state data in accounts. It uses the account address as a lookup path, starting from the root node and proceeding down to the leaf nodes. This lookup capability is not present in binary Merkle trees.

Hyperledger Fabric uses a multi-branch Merkle Bucket tree to compute the State Root. The leaf nodes store key-value type state data sets, and when calculating the root hash, buckets that have not changed can be skipped, making the process very fast.

\subsubsection{Blocks and Nodes}
A block is the basic unit of a blockchain and is divided into a block header and a block body. The transaction list, which consists of all parts of the Merkle tree except for the root hash, is stored in the block body. The block header contains macro information such as the block version number, timestamp, a hash pointer to the previous block (hash of the last header block), the Merkle root hash, mining difficulty target (target), and a nonce. In Bitcoin, nodes are categorized into full nodes and lightweight nodes. Full nodes store the entire block content, while lightweight nodes only store the block header content. The differences between them are shown in Table 2.

\begin{table}[h]
\centering
\caption{Full Node and Light Node}
\begin{tabular}{ p{6.6cm} p{6.6cm}}
\toprule
\textbf{Full Node} & \textbf{Light Node} \\
\midrule
Always stays online & Does not need to stay online\\

Keeps the complete blockchain information on the local hard drive & Does not store the entire blockchain information, only the headers of each block\\

Maintains UTXO set for fast verification of valid transactions & Does not store all transactions, only those relevant to its own transactions\\

Monitors transaction data from specific networks to verify the legality of the transactions  & Cannot verify the legality of many transactions, only can verify the legality of transactions related to itself\\

Cannot verify the legality of many transactions, only can verify the legality of transactions related to itself & Cannot detect the correctness of blocks published on the network\\

Needs to listen to blocks mined in specific regions to verify its legality & Can verify the difficulty of mining\\

Can perform a search, selecting the longest chain, but cannot verify if this longest chain is valid & Can only detect the longest chain, but cannot verify if this longest chain is valid\\

\bottomrule
\end{tabular}%
\end{table}

\subsubsection{Blockchain Structure}
Each node in the chain is linked to another node through the previous block's hash pointer, forming a chain-like structure. If any block is tampered with, it will cause a chain reaction, altering all subsequent blocks.

\subsection{Data Models}
\subsubsection{Transaction-Based Model}
Bitcoin uses a transaction-based data model (transaction-based ledger), where each transaction consists of two parts: inputs and outputs. The inputs indicate the source of the transaction, and the outputs specify the destination. As a result, all transactions are linked together and can be traced back to the Coinbase source created through mining\cite{013} and the unspent coins. The transaction inputs include the Bitcoin holder's signature, the hash of the previous transaction, and the output index. The outputs include the transfer amount and the recipient's public key hash, which is their address. The signature and public key hash ensure non-forgeability. Bitcoin full nodes maintain a UTXO (Unspent Transaction Output) data structure, which helps confirm whether a Bitcoin has been spent, effectively preventing double-spending attacks\cite{7}. The balance of a Bitcoin account\cite{014} is the sum of all unspent Bitcoins under a specific address, which can be more complex to verify.

\subsubsection{Account-Based Model}
Ethereum and Hyperledger Fabric, with their rich functionalities, use an account-based model (account-based ledger), allowing for quick queries of current balances or states \cite{8}. Ethereum accounts are divided into external and contract accounts, representing Ether balances and smart contracts, respectively. Unlike the transaction-based model, the state variables and account balances in smart contracts are state data. Ethereum transaction data includes the recipient's address, the transfer amount, the per-unit gas price, the maximum allowed gas consumption, transaction count, message data for invoking smart contracts\cite{015}, and the sender's signature.

\subsection{Data Storage}
Bitcoin, Ethereum, and Hyperledger Fabric all use Level DB databases to store index data. Level DB is a lightweight, single-node database that does not require installation or deployment, offers high write performance, and can meet the system’s needs for large amounts of hash-based key-value retrievals. However, an architecture based on Level DB cannot meet the business requirements of enterprise-level applications. Hyperledger Fabric 1.0 provides a plug-in data access mechanism, supporting both Level DB and the distributed CouchDB database\cite{9}.

\section{ Network Layer}
Information exchange takes place at the network layer, with blockchain platforms selecting the P2P protocol as the network transmission protocol, which can tolerate single points\cite{10} of failure, thus achieving decentralization. All nodes in the network are equal and autonomous, and they can freely join or exit the network. Any two nodes can directly transact with each other\cite{11}. In a blockchain network, all broadcasted data is constantly monitored by nodes. When a neighboring node sends new data, the data’s validity is first verified. If valid, it is processed; if not, it is discarded.

Bitcoin uses an unstructured networking approach based on the TCP protocol, which facilitates firewall traversal with a randomized routing table. If a node wants to join the blockchain network, it must first contact a seed node, which will then inform the new node of other nodes in the network. Nodes use multicast to transmit data, initially based on the Gossip protocol\cite{12}, but it was later implemented using the Diffusion protocol\cite{13} to improve resistance to anonymous analysis. In terms of data transmission, Bitcoin uses the anonymous communication network Tor for data delivery, with multi-layer encryption on the path protecting the identity of the endpoints. Bitcoin's design principle is simplicity and robustness rather than efficiency.
 
Ethereum's underlying peer-to-peer network protocol cluster is called DEVP2P. In addition to meeting the blockchain network’s functionality, it also meets the needs of any associated Ethereum applications. A node’s public key is used as an identifier, and the Kademlia algorithm is applied to compute the XOR distance between nodes, enabling structured networking. Data propagation is achieved through Gossip, which is multi-point broadcast. For data transmission protection, nodes use Elliptic Curve Integrated Encryption Scheme (ECIES) to generate public and private keys for data encryption.
Hyperledger Fabric is based on the HTTP/2 protocol and builds node clusters by organization. The network uses a hybrid peer-to-peer structure. An organization includes both regular nodes and anchor nodes, which are responsible for routing messages within the organization and across organizations, respectively. Hyperledger Fabric uses Gossip for network initialization, and nodes periodically broadcast their survival status, with other nodes updating their routing tables accordingly. Unlike Bitcoin and Ethereum, Hyperledger Fabric is a permissioned blockchain requiring stricter network layer security mechanisms.

\section{Consensus Layer}
In a decentralized network, nodes lack trust in one another. Achieving consensus and ensuring all nodes act in unison to make the correct decisions is known as the Byzantine Generals Problem\cite{14}.

Bitcoin uses the PoW (Proof of Work) algorithm for node consensus, with the threshold set to produce one block every 10 minutes. A reward and penalty mechanism ensures the sustainability of consensus, mainly through transaction fees, mining rewards, and mining pool distribution strategies. Ethereum uses the PoW consensus, with the threshold set to produce one block every 15 seconds, and plans to adopt the PoS (Proof of Stake) or Casper consensus protocol in the future. The lower computational difficulty may lead to frequent branch chain formations, so Ethereum uses its unique reward and penalty mechanism, the GHOST protocol (Greedy Heaviest-Observed-Sub-Tree)\cite{15}, to incentivize miners’ consensus participation. Hyperledger Fabric initially used the PBFT (Practical Byzantine Fault Tolerance) consensus protocol. Later, to improve transaction throughput and reduce security, the consensus process was decomposed into two services: ordering and validating. As a permissioned chain, Hyperledger Fabric participants have transparent identities and shared intentions, so there is no possibility of node idleness or malicious attacks, eliminating the need for a reward and penalty mechanism. This section will analyze the PoW, PoS, and PBFT algorithms.

\subsection{Proof of Work}

Voting-based consensus algorithms are not suitable for public chains. On one hand, attackers can create many malicious nodes to increase their voting power and launch Sybil attacks\cite{16}; on the other hand, since nodes in a public chain can freely join or leave, not all nodes can guarantee timely participation in voting. Bitcoin and Ethereum use the Proof of Work (PoW) mechanism, where nodes compete for the right to record blocks based on computational power, effectively ensuring data consistency and security. PoW, also known as mining, originates from the work of Dwork \cite{17} and is the first-generation consensus mechanism. All nodes can participate in mining, competing to solve mathematical problems in the blockchain. The block's accounting rights are assigned to the first miner who solves the problem, and the miner is responsible for packaging the transaction into a new block and publishing it. The miner also receives a block reward. In simple terms, you get rewards for the work you do. Nodes that do not receive the accounting rights will verify the content of the new block after receiving it. Only valid data is accepted and added to the local blockchain, and new blocks are built on top. When a new transaction occurs, it is broadcasted to the network. Mining nodes collect all transaction data that has occurred since the last block was published, calculate the Merkle root of the transactions within a specific period, and increment a random number in the block header starting from zero until they find a nonce that satisfies the condition $H(block header) \le  target$. Bitcoin uses the SHA-256 hash algorithm for this calculation, while Ethereum uses the Ethash algorithm.

Mining is an ongoing process of attempting various nonces to solve a puzzle. Each attempt can be viewed as a Bernoulli trial, where each trial is random. A Bernoulli trial has the property of being memoryless, ensuring the fairness of mining. Stronger miners do not have a proportional advantage, and previous mining efforts do not increase the chance of success. Bitcoin’s block production time follows an exponential distribution\cite{18}, with an average block time of 10 minutes for the system. As the total computational power of the system increases, the block time becomes shorter, which can lead to system forks, compromising system security. To maintain a consistent average block time, Bitcoin adjusts the difficulty level every 2016 blocks. The mining difficulty is inversely proportional to the target threshold, as shown in formula (1), where $difficulty\_1\_target$ represents the target threshold corresponding to a mining difficulty of 1.
\begin{equation}
     difficulty= \frac{difficulty\_1\_target}{target} 
     \label{eq1}
\end{equation}

The system adjusts mining difficulty based on formula (2).

\begin{equation}
     target=target\times  \frac{actual\_time}{expected\_time} 
     \label{eq2}
\end{equation}

The block reward is halved every 4 years, becoming smaller over time. However, the incentive for mining intensifies due to the skyrocketing price of Bitcoin. When the block reward decreases to nearly zero, transaction fees will become the primary incentive for mining.

Under the PoW mechanism, the difficulty and cost of an attack are extremely high. If an attacker wants to alter a specific block, they must recalculate the nonce of that block and all subsequent blocks. To accomplish this, the attacker’s computational speed must exceed that of the main chain. Only by controlling more than 51\% of the network's total computational power can the attack succeed\cite{19}. Unlike other consensus algorithms, the PoW mechanism integrates economic incentives, which attracts more nodes to actively participate in mining and encourages nodes to maintain honesty, effectively improving the network's security and reliability. However, PoW consumes significant electricity, which contradicts the human pursuit of energy conservation, cleanliness, and environmental sustainability. Furthermore, over time, the providers of computational power are no longer just individual computers. Users have evolved from personal mining to large mining pools and data centers, leading to more concentrated computational power, which goes against the decentralization principle and gradually threatens network security. Bitcoin's block reward is halved every 4 years, and once mining costs exceed the mining rewards, people's enthusiasm for mining will decrease, leading to a reduction in computational power and triggering security issues.
\subsection{Proof of Stake}
The Proof of Stake (PoS) mechanism first appeared in Peercoin \cite{20}. This mechanism binds difficulty with coin age, where a node must hold coins for a certain period to mine a block. Coin age is the product of the number of coins held and the length of time those coins are held. The mining difficulty decreases proportionally with coin age, thus accelerating block production speed. The right to mine is distributed based on the amount and duration of staked Ether.

In essence, PoS, like PoW, still relies on hash computations for mining. Under the PoS mechanism, only nodes holding more than 50\% of the system’s tokens can initiate a 51\% attack. However, the rewards from such an attack are less than those earned by remaining an honest node, thus enhancing the system's security. PoS substitutes external computational power with stake-based power, solving the resource-wasting issue of PoW. However, when the network environment is poor, forks may occur, which affects the blockchain’s integrity.

Ethereum, based on the PoS mechanism, introduced the Casper consensus \cite{21}, a Security-Deposit Based Economic Consensus Protocol. In this protocol, nodes act as validating nodes by locking a deposit. They must purchase Ether and stake it into Ethereum to participate in consensus, essentially betting on the consensus. Only those who have staked Ether can participate in block production and consensus formation. Consensus results are formed based on the betting patterns of the validating nodes. These nodes must predict which block other nodes will bet on, and if they bet correctly, they retrieve their stake along with transaction fees. If consensus is not quickly reached, only part of the deposit is returned. Over several rounds, the betting distribution of validating nodes converges. Thus, transactions are essentially bets on certain blocks. Once a block is confirmed, the validating nodes who bet on that block receive rewards, while nodes that bet on other blocks are fined by losing part of their staked Ether.

Many early PoS algorithms only considered rewarding block creation without implementing penalties, which led to undesirable outcomes. In the case of multiple competing blockchains, validating nodes could create blocks on each chain to ensure rewards. If all participants are purely profit-driven, consensus may not be achieved even in the absence of attackers. Casper includes penalty mechanisms where dishonest miners lose their entire stake, and their rights are revoked. This resolves the issue in PoS protocols of low cost for malicious actions, often referred to as the "no-stakes problem." Additionally, if validating nodes change their bets too significantly—such as first betting on one block with a high probability of winning, then switching to another block with a high probability of winning—they will be severely punished. This rule ensures that validating nodes only bet heavily on blocks they are very confident others will also bet on. This mechanism prevents a situation where betting first converges on one result, only to later shift to another. Casper does not require additional power consumption for mining, reduces block times to 4 seconds, and is more resource-efficient and faster than PoW.

\subsection{ Practical Byzantine Fault Tolerance}
The PoW mechanism, based on proof-of-work, relies on computational power competition to ensure data consistency and security in blockchain networks. While it is suitable for public chains with freely joining and exiting nodes, it consumes a significant amount of computational resources and energy, making it unsuitable for consortium blockchains used in enterprise applications. Hyperledger Fabric adopts the Practical Byzantine Fault Tolerance (PBFT) algorithm \cite{22}, which is based on a voting mechanism. PBFT increases the system’s fault tolerance to around 33\%, ensuring data consistency and security as long as honest nodes exceed two-thirds of the total nodes. PBFT is computation-based and does not offer token rewards. The system selects a primary node through rotation or a random algorithm. As long as the primary node remains unchanged, this is referred to as a "view." In this view, when a client initiates a transaction, it sends a request to the primary node, which then broadcasts the message to all backup nodes. The backup nodes validate the message, and if validated, they send a confirmation. When confirmation messages exceed 2f+1, the result is returned to the client and written into the blockchain. The process consists of three phases: the pre-prepare phase, the prepare phase, and the commit phase. Each block is generated by a unique primary node, eliminating the risk of forks. However, when more than one-third of the accounting nodes stop working, the system can no longer function. Additionally, in large networks with many nodes, validating messages creates significant network overhead. Therefore, PBFT is better suited for private chains or consortium chains with fewer nodes, rather than public chains with large, dynamic node networks.
\subsection{Comparison of Consensus Algorithms}
Currently, no consensus algorithm offers optimal performance in all aspects. The choice depends on specific requirements and trade-offs. Table 3 compares PoW, PoS, and PBFT. To ensure the security and decentralization of blockchain systems, performance must be continually improved to accommodate large-scale applications, while also motivating users to actively participate in the consensus process.

\begin{table}[h!]
\centering
\caption{Comparison of Consensus Algorithms}
\begin{tabular}{m{1.8cm}p{2.5cm}p{2.5cm}p{2.5cm}p{2.5cm}}
\toprule
\textbf{Indicator} & \textbf{Performance Efficiency} & \textbf{Degree of Decentralization} & \textbf{Fault Tolerance Rate} & \textbf{Resource Consumption} \\
\midrule
PoW  & Low & High & 50\% & High \\
PoS  & Relatively High & High & 50\% & Low \\
PBFT & High & Low & 33\% & Low \\
\bottomrule
\end{tabular}
\end{table}

\section{Smart Contract}
A smart contract\cite{23} is a set of digitally defined commitments. On the blockchain, it is an executable program code\cite{24} designed to automatically enforce the terms and conditions between two untrusted parties. Blockchain-based smart contracts have interfaces to receive and respond to external messages, with a greater focus on transactions. Smart contracts neither generate nor modify data; they serve as modules for transaction processing and state recording, enabling the automatic execution of contract terms under predefined conditions, achieving the goal of "code as law." Smart contracts can be classified into three categories based on their execution environment and programming language: script-based, Turing-complete, and verifiable smart contracts.

The Bitcoin system uses script-based encoding to facilitate basic transactions of digital currency. These scripts are not Turing-complete programming languages; they are simply a set of limited, type-specific stack instructions. There is no strict implementation of smart contracts in the Bitcoin platform, which can only be considered as an early prototype of smart contracts. Ethereum, in its whitepaper, was the first to introduce the application of smart contracts to the blockchain, effectively reviving the concept of smart contracts. Hyperledger Fabric supports a pluggable consensus mechanism, which provides developers with convenient options for selecting appropriate consensus algorithms.

\subsection{Programming Languages}
Ethereum has custom-designed Turing-complete scripting languages such as Solidity and Serpent, which enhance contract logic functionality and reduce the complexity of contract creation, but also introduce security risks. After a smart contract is written, it is compiled into Ethereum Virtual Machine (EVM) bytecode by a compiler and uploaded to the blockchain by the client. It is then executed in the Ethereum Virtual Machine (EVM) by miners. Michael Coblenz and others have identified numerous bugs in smart contracts designed with Solidity and developed Obsidian\cite{25} as a solution. Hyperledger Fabric allows for the development of smart contracts based on advanced programming languages such as Go and Java. These high-level languages are not only Turing-complete but also feature mature compilation techniques, which help reduce the learning curve for contract developers.

\subsection{Execution Environment}
Smart contracts must run in an isolated sandbox environment because they are human-written, and therefore, contain numerous potential vulnerabilities. They cannot run directly in the known environment of blockchain nodes. In a sandbox environment, contracts are effectively isolated from both the host and other contracts, improving security. The Ethereum Virtual Machine (EVM) \cite{27}is a custom-built sandbox for Ethereum. It does not have a network interface, compiled bytecode cannot access the host machine, and the inter-contract calls are highly restricted. For a single smart contract, multiple instances often need to run simultaneously across several Ethereum Virtual Machines to ensure data consistency and high fault tolerance across the blockchain, though this also limits the overall capacity of the network. Hyperledger Fabric, on the other hand, uses lightweight Docker containers as sandboxes, providing isolated Linux environments. Contracts in Docker containers still have access to the internet.

\subsection{Operating Principle}

\begin{figure}
    \centering
    \includegraphics[width=0.96\linewidth]{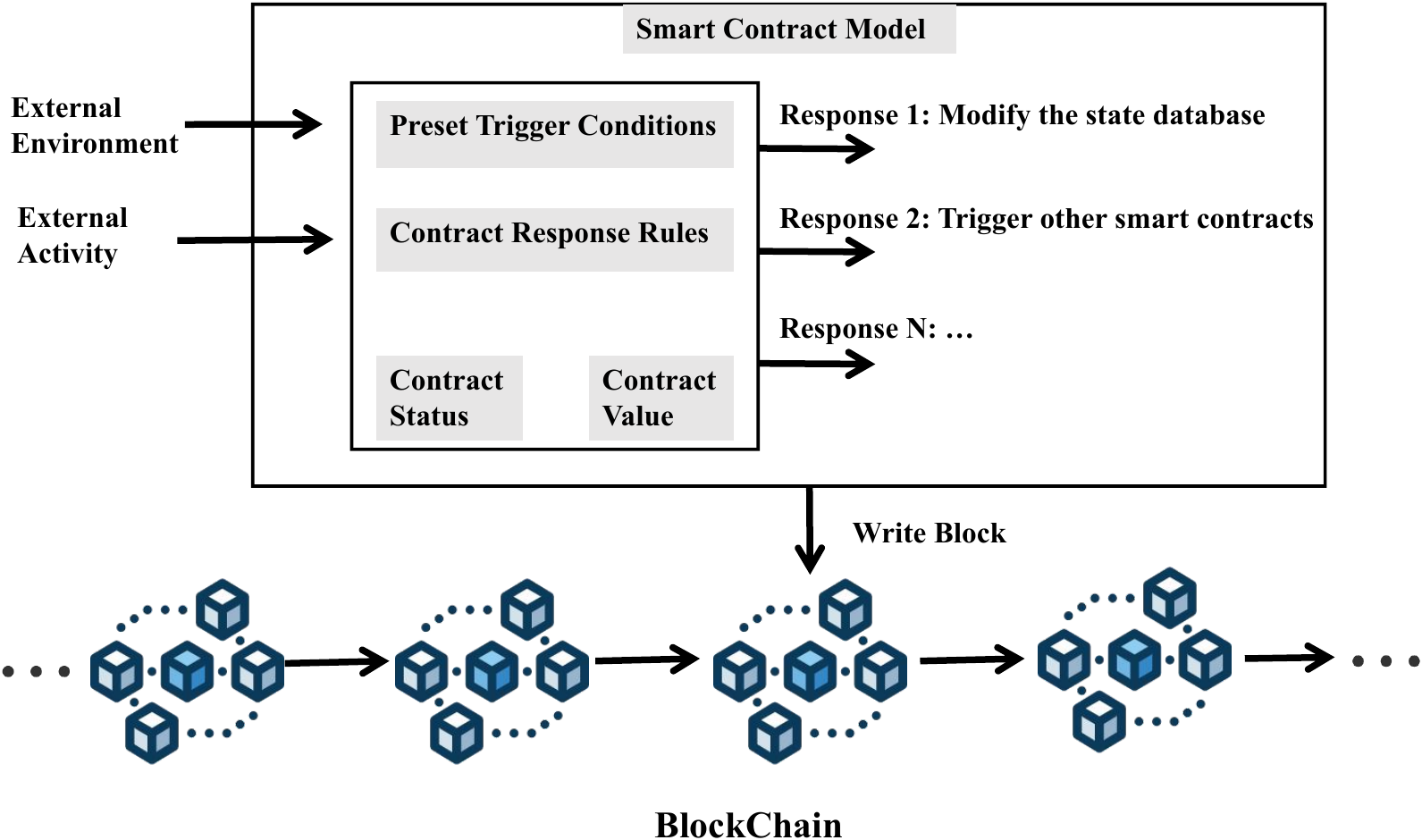}
    \caption{Operational Principle of Smart Contracts}
    \label{fig5}
\end{figure}

Currently, smart contracts are simply a set of fixed rules implementing an IF-THEN program structure. They are not truly intelligent, but can monitor the state data of the blockchain in real time. As shown in Figure \ref{fig5}, when the external environment changes to match a predefined state, the contract is triggered, and the account’s state is modified. If an external application wants to invoke a smart contract to modify a particular element, it must first obtain consensus across the entire network. Only then will the blockchain record this modification and save the result to the state database.

Ethereum is the most widely used Turing-complete blockchain platform\cite{28}. It uses accounts to record system states directly, allowing developers to write smart contracts that include ownership, transaction formats, and state transition functionalities, following random rules. Anyone can participate in the Ethereum blockchain network\cite{29} on their machine. Ethereum supports data transfer between different accounts to enable more complex logic. The main components of an Ethereum smart contract are functions, events, and state variables written stably\cite{30}. Ethereum accounts are of two types: Contract Accounts and Externally Owned Accounts (EOAs). Contract accounts are used to store the executed smart contract code and can only be activated by externally owned accounts. Externally owned accounts are owned by Ether holders and correspond to a specific public key. Accounts include fields such as nonce, balance, storageRoot, and codeHash, which are controlled by individuals. Ethereum smart contract accounts\cite{31} consist of executable code, contract addresses, state, and virtual currency balances. When a contract account is invoked, the smart contract automatically executes in the virtual machine. During execution, it consumes gas (fuel), and when the gas runs out, the contract execution halts immediately and reverts the state. The consumed gas is not refunded, which effectively prevents spam transactions and infinite loop attacks. Ethereum's account address data is built-in, making it more suitable for payment applications based on digital currencies. In Ethereum, a transaction refers to message data transferred from one account to another. Ethereum uses transactions as the smallest execution unit. Similar to Bitcoin, users must pay a transaction fee when sending transactions, using Ether as payment and consumption. Currently, the Ethereum network supports a transaction rate higher than Bitcoin's (up to several dozen transactions per second). The currency in the Ethereum network is Ether, mainly used to purchase gas to maintain the cost of running smart contracts. Ether, like Bitcoin, can be mined, and miners who generate new blocks are rewarded with five Ether and the transaction fees contained in the block. Users can also directly purchase Ether from others. Currently, more than ten million Ether can be generated annually through mining, with a market price exceeding 300 USD per Ether. Gas controls the maximum execution instructions for a given transaction and can be exchanged for Ether. It is important to note that the price of Ether fluctuates, but the fuel cost for running a particular smart contract can be fixed by adjusting parameters like the Gas price.

In Hyperledger Fabric, nodes start new containers based on contracts packaged into Docker images. These containers initialize according to the rules in the smart contract and then wait for invocation. Hyperledger Fabric's smart contracts are called Chaincode\cite{32} and are the only way to interact with the blockchain and generate transactions. Chaincode can be seen as managing transactions, and it controls how to package smart contracts for deployment. Chaincode enforces rules to read or modify key-value pairs or other state database information, executing on the current state of the ledger and starting through transaction proposals. Chaincode executes a set of key-value writes that can be submitted to the network and applied to the ledger across all peers. By granting appropriate permissions, chaincode status in one network can be accessed via the same network's chaincode Go API. Writing a contract essentially means implementing the initialize, modify, and query methods in the chaincode interface. Multiple smart contracts can be defined within the same main code, and when a chaincode is deployed, all the smart contracts within it can be used by applications. At the simplest level, the blockchain immutably records transactions updated in the ledger. Smart contracts programmatically access two distinct parts of the ledger: the blockchain, which immutably records all transaction histories, and the world state, which stores the current values of these states as cached objects representing the current value of typically needed objects. Smart contracts allow for the retrieval and deletion of states in the world state and can query immutable blockchain transaction records. Whether a transaction creates, reads, updates, or deletes business objects in the world state, the blockchain contains immutable records of these changes. Each chaincode is associated with an endorsement policy that applies to all smart contracts defined within it. Important transaction instructions in the blockchain are announced in the smart contract chain, and significant transactions must be validated via the blockchain. Each smart contract has an associated endorsement policy, which determines which organizations must approve the transactions generated by the smart contract before they can be marked as valid. If the endorsement policy specifies that multiple organizations must sign off on the transaction, the smart contract must be executed by enough organizations to generate a valid transaction. The endorsement policy distinguishes Hyperledger Fabric from Ethereum and Bitcoin, where any node in the network can generate valid transactions. Hyperledger Fabric more accurately simulates the real world by requiring transactions to be validated by trusted network organizations; the endorsement policy is designed to model these real-world interactions better. Smart contracts run on peer nodes owned by organizations in the blockchain network. The contract accepts a set of transaction proposals and combines program logic to read and write the ledger. Changes to the world state are captured as transaction proposal responses (or just transaction responses), with the read-write set containing the states that were read and the new states to be written when the transaction is validated. Transactions, distributed to all peers in the network, are processed in two phases: first, the endorsement policy checks whether enough organizations have signed the transaction; second, it checks to ensure that the current value of the world state matches the read set of the transaction when signed by the endorsing peers, with no intermediate updates. If the transaction passes both tests, it is marked as valid. All valid and invalid transactions are added to the blockchain's history, but only valid transactions lead to updates to the world state.

\section{Application Layer}
The application layer includes services and applications. After the underlying data and computing tools are integrated, they provide services to upper-layer applications. Current service platforms lack general applicability and are mainly designed to meet the needs of specific services. Bitcoin is primarily used for digital currency transactions, while Ethereum also supports communication with smart contracts through JSONRPC to enable decentralized applications (DApps) built with JavaScript. Hyperledger Fabric is mainly geared toward enterprise-level applications and does not provide digital currencies. Its applications can be built using SDKs in languages such as Go, Java, Python, and Node.js.
\section{Scalability}
With the increasing transaction volume, the blockchain is becoming larger and larger. Each node must store all transaction information and validate it, and the original block size is limited with a fixed block time, which cannot meet the demand for real-time processing of millions of data. For example, Bitcoin can only process around 7 transactions per second \cite{34}. Furthermore, because the capacity of mining pools is limited, miners tend to select transactions with higher fees, which causes delays in smaller transactions. To ensure faster transaction confirmations, the transaction fees paid to miners gradually rise. To resolve the issue of transaction volume saturation and high transaction fees, and to accelerate blockchain application development and expansion, it is necessary to address the bottleneck in improving transaction throughput and scaling the blockchain.

When Bitcoin was first created, the block size was not strictly limited. According to Bitcoin's data structure rules, a block can be as large as 32MB. However, during the initial phase, the average block size was only 1–2KB, far from reaching the block size limit, which caused resource waste and made the system susceptible to Distributed Denial-of-Service (DDoS) attacks. To ensure the system's security and stability, Bitcoin later limited the block size to 1MB. Different user groups have different opinions on scaling. The primary disagreement is between two factions: the core development team, which wants to maintain Bitcoin's small block characteristics, and miners and developers who oppose using centralized solutions like the Lightning Network. To date, many versions of Bitcoin Improvement Proposals (BIPs) have been proposed to address the scaling issue. However, BIPs are only proposals, as their implementation requires changes to the Bitcoin source code, and achieving a scaling solution requires consensus within the entire Bitcoin community. Generally, Bitcoin’s scaling solutions can be categorized into two types: those that directly change the block size and those that improve transaction processing capacity without changing the block size. Regardless of the approach, a difficult consensus process is required, but scaling is a necessary issue for Bitcoin’s sustainable development.

Buterin proposed a solution for sharding transaction processing in the Ethereum 2.0 whitepaper \cite{35}. Sharding involves dividing the network into many sub-networks, where each shard consists of nodes that maintain and execute the same set of smart contracts. Sharding technology addresses issues from a resource-balancing perspective. The system divides the entire network into many shards that are independent of each other, with no interdependencies between them. Each shard maintains its independent sub-chain. Nodes can freely choose which shard to join or which shard to execute transactions on. Once a node joins a shard, it is responsible for storing and processing the transactions on that shard. Nodes work together, and data storage and processing can be parallelized. The overall network processing capacity is no longer constrained by a single node. However, because of sharding, the computational power of the network is distributed, making it easier for an attacker to control 51\% of the network's computational power, meaning that PoW is no longer applicable. To mitigate this, the network uses randomness to randomly select nodes to form shards. This random sampling method prevents malicious nodes from filling up individual shards. Additionally, a consensus protocol, such as PoW, is required to ensure that members within a shard reach an agreement. The randomness in blocks is publicly verifiable, and unified random bits can be extracted.

Hyperledger Fabric 1.0 proposes a multi-channel solution, where the entire network is divided into many logical channels. Nodes can freely join channels and process data on different chains simultaneously, allowing transactions to be executed independently and concurrently, which improves the network's throughput.

\section{Security}
\subsection{Data Security}
In public blockchains like Bitcoin and Ethereum, any node can freely join or leave the network, and any user can participate in transactions. All data is open and transparent, which poses significant security risks.

In the Bitcoin system, for a transaction to occur, the recipient must provide the sender with an address hashed from the public key, and the sender must sign the transaction data. When the hash value in the transaction input matches the hash value in the output of the previous transaction, it confirms that the Bitcoin is indeed owned by the sender, thus preventing double-spending attacks to some extent. The signing and verification process is automated via scripts, without requiring manual calculation. Unlike Bitcoin, which is transaction-based, Ethereum is account-based, and its transaction signing and verification process differs. In Ethereum, the transaction data contains only the sender’s EDCSA signature, and the sender’s public key can be derived from the signature, transaction data, and elliptic curve parameters. The account address is then obtained by performing a SHA3 hash on the public key. This reduces the number of bytes in the transaction, lowering the overhead.

In Hyperledger Fabric, to meet the requirements of a consortium blockchain, unauthorized nodes cannot join the network. Hyperledger Fabric provides membership management services and offers three types of digital certificates: ECert for identity authentication, TCert for transaction signing and verification, and TLSCert for secure communication between system components based on SSL/TLS.

\subsection{Privacy}
All data in blockchain systems is public and transparent, which helps prevent data falsification and tampering, but privacy is not guaranteed. Bitcoin proposed a privacy protection strategy by anonymizing the link between user accounts and transactions. Accounts can generate a large number of public/private key pairs using a seed node and then generate multiple addresses using hash functions. The real-world identity of the user is not directly linked to the account address, achieving a degree of anonymity. However, this anonymity is somewhat illusory, as attackers can analyze transaction records to determine which network a node belongs to. As a result, the account addresses in the network could still be linked to real-world addresses.

To protect blockchain privacy, it is necessary not only to hide transaction details but also to verify the correctness of transactions. Zero-Knowledge Proof (ZKP) technology\cite{36} allows for hiding the sender, receiver, and transaction details. In Ethereum, EliBen and others studied the use of ZK-STARK to enhance scalability and privacy protection.

Hyperledger Fabric 0.6 adopted a single-chain approach, where all users on the chain could access all transaction data, which does not meet the privacy requirements of a consortium blockchain for commercial organizations. Hyperledger Fabric 1.0 adopted a multi-channel approach, establishing separate channels for transaction nodes, storing transaction data within these channels, and preventing users outside the channel from accessing the data, thereby ensuring privacy.

\section{Conclusion}
Without the coordination of a third-party authority, blockchain enables trustworthy data transmission between parties who do not know each other. This plays a crucial role in driving social management and applications in various fields, and the research and development prospects of blockchain are vast. Research into the basic principles and technologies serves as the foundation for all work. This paper provides a comprehensive and comparative analysis of the key technologies in different blockchain systems, from the perspective of the various layered architectures of blockchain networks. It examines their similarities, differences, and advantages and disadvantages. In today's rapidly advancing technology landscape, more innovations will continue to emerge and impact the development of various blockchain systems. The ongoing challenge for all blockchain systems will be how to adapt to technological updates and iterations, secure a place in the internet age, and meet people's demands with higher service experiences and standards.

\bibliographystyle{ACM-Reference-Format}
\bibliography{ref}

@String{JACM = "J. ACM" }

@String{Computing = "Computing" }

@String{Computer = "{IEEE} Computer" }

@String{Springer = "Springer-Verlag" }

@article{1,
  title={A peer-to-peer electronic cash system},
  author={Nakamoto, Satoshi and Bitcoin, A},
  journal={Bitcoin.--URL: https://bitcoin. org/bitcoin. pdf},
  volume={4},
  number={2},
  pages={15},
  year={2008}
}

@article{2,
  title={A next-generation smart contract and decentralized application platform},
  author={Buterin, Vitalik and others},
  journal={white paper},
  volume={3},
  number={37},
  pages={2--1},
  year={2014}
}

@inproceedings{3,
  title={Architecture of the hyperledger blockchain fabric},
  author={Cachin, Christian and others},
  booktitle={Workshop on distributed cryptocurrencies and consensus ledgers},
  volume={310},
  number={4},
  pages={1--4},
  year={2016},
  organization={Chicago, IL}
}

@article{4,
  title={Blockchain: the state of the art and future trends},
  author={Yuan, Yong and Wang, Fei-Yue and others},
  journal={Acta automatica sinica},
  volume={42},
  number={4},
  pages={481--494},
  year={2016}
}

@book{5,
  title={Introduction to modern cryptography: principles and protocols},
  author={Katz, Jonathan and Lindell, Yehuda},
  year={2007},
  publisher={Chapman and hall/CRC}
}

@article{6,
  title={PATRICIA—practical algorithm to retrieve information coded in alphanumeric},
  author={Morrison, Donald R},
  journal={Journal of the ACM (JACM)},
  volume={15},
  number={4},
  pages={514--534},
  year={1968},
  publisher={ACM New York, NY, USA}
}

@inproceedings{7,
  title={Double-spending fast payments in bitcoin},
  author={Karame, Ghassan O and Androulaki, Elli and Capkun, Srdjan},
  booktitle={Proceedings of the 2012 ACM conference on Computer and communications security},
  pages={906--917},
  year={2012}
}

@book{8,
  title={Bitcoin and cryptocurrency technologies: a comprehensive introduction},
  author={Narayanan, Arvind and Bonneau, Joseph and Felten, Edward and Miller, Andrew and Goldfeder, Steven},
  year={2016},
  publisher={Princeton University Press}
}

@book{9,
  title={CouchDB: the definitive guide: time to relax},
  author={Anderson, J Chris and Lehnardt, Jan and Slater, Noah},
  year={2010},
  publisher={" O'Reilly Media, Inc."}
}

@inproceedings{10,
  title={What can database do for peer-to-peer?},
  author={Gribble, Steven D and Halevy, Alon Y and Ives, Zachary G and Rodrig, Maya and Suciu, Dan},
  booktitle={WebDB},
  volume={1},
  pages={31--36},
  year={2001}
}

@book{11,
  title={Mastering Bitcoin: unlocking digital cryptocurrencies},
  author={Antonopoulos, Andreas M},
  year={2014},
  publisher={" O'Reilly Media, Inc."}
}

@inproceedings{12,
  title={Epidemic algorithms for replicated database maintenance},
  author={Demers, Alan and Greene, Dan and Hauser, Carl and Irish, Wes and Larson, John and Shenker, Scott and Sturgis, Howard and Swinehart, Dan and Terry, Doug},
  booktitle={Proceedings of the sixth annual ACM Symposium on Principles of distributed computing},
  pages={1--12},
  year={1987}
}

@article{13,
  title={Consensus in the presence of partial synchrony},
  author={Dwork, Cynthia and Lynch, Nancy and Stockmeyer, Larry},
  journal={Journal of the ACM (JACM)},
  volume={35},
  number={2},
  pages={288--323},
  year={1988},
  publisher={ACM New York, NY, USA}
}

@incollection{14,
  title={The Byzantine generals problem},
  author={Lamport, Leslie and Shostak, Robert and Pease, Marshall},
  booktitle={Concurrency: the works of leslie lamport},
  pages={203--226},
  year={2019}
}

@article{15,
  title={Accelerating bitcoin's transaction processing. fast money grows on trees, not chains},
  author={Sompolinsky, Yonatan and Zohar, Aviv},
  journal={Cryptology ePrint Archive},
  year={2013}
}

@inproceedings{16,
  title={The sybil attack},
  author={Douceur, John R},
  booktitle={International workshop on peer-to-peer systems},
  pages={251--260},
  year={2002},
  organization={Springer}
}

@inproceedings{17,
  title={Pricing via processing or combatting junk mail},
  author={Dwork, Cynthia and Naor, Moni},
  booktitle={Annual international cryptology conference},
  pages={139--147},
  year={1992},
  organization={Springer}
}

@article{18,
  title={Proof of stake versus proof of work},
  author={BitFury, G},
  journal={White paper, Sep},
  volume={810},
  year={2015}
}

@techreport{19,
  title={Exposing computationally-challenged Byzantine impostors},
  author={Aspnes, James and Jackson, Collin and Krishnamurthy, Arvind},
  year={2005},
  institution={Technical Report YALEU/DCS/TR-1332, Yale University Department of Computer}
}

@article{20,
  title={Ppcoin: Peer-to-peer crypto-currency with proof-of-stake},
  author={King, Sunny and Nadal, Scott},
  journal={self-published paper, August},
  volume={19},
  number={1},
  year={2012}
}

@article{21,
  title={Proof of stake with casper the friendly finality gadget protocol for fair validation consensus in ethereum},
  author={Jain, Akshita and Arora, Sherif and Shukla, Yashashwita and Patil, T and Sawant-Patil, S},
  journal={International Journal of Scientific Research in Computer Science, Engineering and Information Technology},
  volume={3},
  number={3},
  pages={291--298},
  year={2018}
}

@inproceedings{22,
  title={Blockbench: A framework for analyzing private blockchains},
  author={Dinh, Tien Tuan Anh and Wang, Ji and Chen, Gang and Liu, Rui and Ooi, Beng Chin and Tan, Kian-Lee},
  booktitle={Proceedings of the 2017 ACM international conference on management of data},
  pages={1085--1100},
  year={2017}
}

@article{23,
  title={Formalizing and securing relationships on public networks},
  author={Szabo, Nick},
  journal={First monday},
  year={1997}
}

@article{24,
  title={Smart contract templates: foundations, design landscape and research directions},
  author={Clack, Christopher D and Bakshi, Vikram A and Braine, Lee},
  journal={arXiv preprint arXiv:1608.00771},
  year={2016}
}

@inproceedings{25,
  title={Obsidian: a safer blockchain programming language},
  author={Coblenz, Michael},
  booktitle={2017 IEEE/ACM 39th international conference on software engineering companion (ICSE-C)},
  pages={97--99},
  year={2017},
  organization={IEEE}
}

@book{27,
  title={Introducing Ethereum and solidity},
  author={Dannen, Chris and others},
  volume={1},
  year={2017},
  publisher={Springer}
}

@book{28,
  title={Advanced applications of blockchain technology},
  author={Kim, Shiho and Deka, Ganesh Chandra},
  volume={60},
  year={2020},
  publisher={Springer}
}

@incollection{29,
  title={Blockchain technology for decentralized autonomous organizations},
  author={Singh, Madhusudan and Kim, Shiho},
  booktitle={Advances in computers},
  volume={115},
  pages={115--140},
  year={2019},
  publisher={Elsevier}
}

@inproceedings{30,
  title={From institutions to code: Towards automated generation of smart contracts},
  author={Frantz, Christopher K and Nowostawski, Mariusz},
  booktitle={IEEE 1st International Workshops on Foundations and Applications of Self Systems},
  pages={210--215},
  year={2016},
  organization={IEEE}
}

@inproceedings{31,
  title={Visual emulation for Ethereum's virtual machine},
  author={Norvill, Robert and Pontiveros, Beltran Borja Fiz and State, Radu and Cullen, Andrea},
  booktitle={NOMS IEEE/IFIP Network Operations and Management Symposium},
  pages={1--4},
  year={2018},
  organization={IEEE}
}

@misc{32,
  title={Development guide of blockchain},
  author={Qing-Chun, Shentu},
  year={2017},
  publisher={Beijing: China Machine Press}
}

@book{34,
  title={Distributed ledger technology: The science of the blockchain},
  author={Wattenhofer, Roger},
  year={2017},
  publisher={CreateSpace Independent Publishing Platform}
}

@article{35,
  title={Ethereum: platform review},
  author={Buterin, Vitalik},
  journal={Opportunities and challenges for private and consortium blockchains},
  volume={45},
  pages={1--45},
  year={2016}
}

@article{36,
  title={zerocash decentralized anonymous payments from bitcoin},
  author={Apostolaki, Maria and Zohar, Aviv and Vanbever, Laurent},
  year={2017}
}

@inproceedings{01,
  title={Cobra: interaction-aware bytecode-level vulnerability detector for smart contracts},
  author={Li, Wenkai and Li, Xiaoqi and Li, Zongwei and Zhang, Yuqing},
  booktitle={Proceedings of the 39th IEEE/ACM International Conference on Automated Software Engineering},
  pages={1358--1369},
  year={2024}
}

@inproceedings{02,
  title={Characterizing the Solana NFT ecosystem},
  author={Kong, Dechao and Li, Xiaoqi and Li, Wenkai},
  booktitle={Companion Proceedings of the ACM Web Conference},
  pages={766--769},
  year={2024}
}

@inproceedings{03,
  title={Unveiling wash trading in popular NFT markets},
  author={Niu, Yuanzheng and Li, Xiaoqi and Peng, Hongli and Li, Wenkai},
  booktitle={Companion Proceedings of the ACM Web Conference},
  pages={730--733},
  year={2024}
}

@inproceedings{06,
  title={StateGuard: Detecting State Derailment Defects in Decentralized Exchange Smart Contract},
  author={Li, Zongwei and Li, Wenkai and Li, Xiaoqi and Zhang, Yuqing},
  booktitle={Companion Proceedings of the ACM Web Conference},
  pages={810--813},
  year={2024}
}

@article{08,
  title={SCLA: Automated Smart Contract Summarization via LLMs and Semantic Augmentation},
  author={Mao, Yingjie and Li, Xiaoqi and Li, Wenkai and Wang, Xin and Xie, Lei},
  journal={arXiv preprint arXiv:2402.04863},
  year={2024}
}

@inproceedings{013,
  title={GasTrace: Detecting Sandwich Attack Malicious Accounts in Ethereum},
  author={Liu, Zekai and Li, Xiaoqi and Peng, Hongli and Li, Wenkai},
  booktitle={IEEE International Conference on Web Services (ICWS)},
  pages={1409--1411},
  year={2024},
  organization={IEEE}
}

@article{014,
  title={SCALM: Detecting Bad Practices in Smart Contracts Through LLMs},
  author={Li, Zongwei and Li, Xiaoqi and Li, Wenkai and Wang, Xin},
  journal={arXiv preprint arXiv:2502.04347},
  year={2025}
}

@inproceedings{015,
  title={Detecting Malicious Accounts in Web3 through Transaction Graph},
  author={Li, Wenkai and Liu, Zhijie and Li, Xiaoqi and Nie, Sen},
  booktitle={Proceedings of the 39th IEEE/ACM International Conference on Automated Software Engineering},
  pages={2482--2483},
  year={2024}
}
\appendix

\end{document}